\documentclass[journal=aamick,manuscript=article]{achemso}

\usepackage[usenames,dvipsnames,table]{xcolor}
\usepackage[version=3]{mhchem} %
\usepackage{miller}
\usepackage{subcaption}
\usepackage{multirow}
\usepackage{color}
\usepackage{textgreek}

\usepackage{amsmath,amssymb}
\graphicspath{{Figures/}}

\usepackage{pdfpages}

\usepackage[%
  breaklinks,             %
  bookmarks = false, %
  pdfpagemode   = UseNone,%
  pdfstartview  = FitH,     %
  pdfstartpage  = 1,      %
  colorlinks    = true,   %
]{hyperref}

\newcommand\myshade{85}
\colorlet{mylinkcolor}{YellowOrange}
\colorlet{myurlcolor}{Aquamarine}
\colorlet{mycitecolor}{violet}

\hypersetup{
  linkcolor  = mylinkcolor!\myshade!black,
  citecolor  = mycitecolor!\myshade!black,
  urlcolor   = myurlcolor!\myshade!black,
  colorlinks = true,
}



\author{Md Salman Rabbi Limon}
\affiliation{Department of Mechanical Engineering, Texas Tech University, Lubbock, Texas 79409, USA}
\author{Zeeshan Ahmad}
\affiliation{Department of Mechanical Engineering, Texas Tech University, Lubbock, Texas 79409, USA}
\email{zeeahmad@ttu.edu}

\title{Heterogeneity in Point Defect Distribution and Mobility in Solid Ion Conductors}


\begin{document}

\begin{abstract}
Alkali metal anodes paired with solid ion conductors offer promising avenues for enhancing battery energy density and safety. To facilitate rapid ion transport crucial for fast charging and discharging, it is essential to understand point defects within these conductors. In this study, we investigate the heterogeneity of defect distribution in Li$_3$OCl solid ion conductor, quantifying the defect formation energy (DFE) of lithium vacancies and interstitials as a function of distance from the surface through first-principles simulations. Our results reveal that the surface DFE is consistently lower than bulk except for one surface termination, indicating significant defect aggregation at surfaces. This difference can cause the defect density to be up to 14 orders of magnitude higher at surfaces compared to the bulk. Moreover, we unveil the transition in DFE when moving from the surface to the bulk through the DFE function, which exhibits an exponentially decaying relationship. Incorporating this exponential trend, we develop a revised model for the average behavior of defects that offers a more accurate description of the influence of grain size. Surface effects dominate for grain sizes $\lesssim$ 1 μm, highlighting the importance of surface defect engineering and the DFE function for accurately capturing ion transport in devices. We further explore the kinetics of defect redistribution by calculating the migration barriers for defect movement between bulk and surfaces. We find a highly asymmetric energy landscape for the lithium vacancies, exhibiting lower migration barriers for movement towards the surface compared to the bulk, while interstitial defects exhibit comparable kinetics between surface and bulk regions. These insights underscore the importance of considering both thermodynamic and kinetic factors in the design of solid ion conductors.

\end{abstract}

\section{Introduction}

Point defects have a significant beneficial or detrimental impact on the electronic, transport, optical, magnetic, and thermal properties of materials~\cite{freysoldt2014first}. 
In semiconductors, the concentration of defects determines the p or n doping, thereby controlling the electrical conductivity~\cite{yuFundamentalsSemiconductorsPhysics2010}. 
Defects are key enablers of ion transport in solid ion conductors that have applications in devices such as solid-state batteries, solid-oxide fuel cells, gas sensors, etc. A high concentration of defects coupled with a low migration barrier promotes fast ion transport~\cite{wangDesignPrinciplesSolidstate2015,panGeneralMethodPredict2015}.

Recently, solid-state batteries, which use a solid ion conductor instead of a liquid electrolyte, have emerged as a promising solution for the electrification of road transport and aviation. Solid-state batteries offer the possibility of enabling alkali metal anodes such as lithium that may provide approximately 50\% higher energy density than the current state of the art~\cite{albertusStatusChallengesEnabling2018,pasta2020RoadmapSolidstate2020,janek2016solid}. Further, solid ion conductors offer better voltage windows, mechanical stability, and enhanced safety compared to flammable liquid electrolytes. 
For the operation of these batteries at fast charge and discharge rates, rapid ion transport through the solid ion conductor is essential. Therefore, understanding the properties of point defects such as lithium vacancies and interstitials that are the building blocks of ion transport within the solid ion conductor is crucial for optimizing battery performance.

Most efforts on developing solid ion conductors have focused on lowering the migration barrier of the hopping ion in the bulk crystal~\cite{wangDesignPrinciplesSolidstate2015,heOriginFastIon2017,linDesignCationTransport2020a,sendek2017holistic,ahmadMachineLearningEnabled2018}. These studies have generated structural design principles for fast ion conduction such as the existence of a body-centered-cubic lattice framework of the anion~\cite{wangDesignPrinciplesSolidstate2015}, availability of a concerted migration path~\cite{heOriginFastIon2017}, and presence of lattice frustration and disorder~\cite{wangFrustrationSuperIonicConductors2023}. However, the surfaces and interfaces of solid ion conductors in solid-state batteries have been found to contribute the most to the potential drops and impedance~\cite{masudaInternalPotentialMapping2017,fullerSpatiallyResolvedPotential2021} and may act as the bottleneck for transport in devices. Moreover, grain boundaries in polycrystalline solid ion conductors introduce additional hindrances to ion transport.
The large disparity in bulk and surface properties~\cite{ahmadInterfacesSolidElectrolyte2021} underscores the importance of going beyond bulk to account for the impact of these factors on performance.
At surfaces occurring near grain boundaries and electrode-electrolyte interfaces in solid-state batteries, the properties of point defects may deviate significantly from the bulk due to broken symmetry, undercoordinated atoms, structural changes, and the presence of decomposition products~\cite{ahmadUnderstandingEffectLead2022}.

One implication of the heterogeneity in the formation energy of point defects is the observed segregation of defects and dopants in mixed conducting oxides~\cite{leeAtomisticSimulationsSurface2010}. This heterogeneity leads to enrichment of dopants/defects in the surface/core region and depletion in the neighboring region called the space charge zone~\cite{Maier1995ionic}. Ionic transport is severely limited in the space charge zone due to ion depletion.  This presents another scenario where the bulk ion transport is not an accurate descriptor for the device performance and the variation in the DFE from bulk to the surface plays a critical role. 
The role of space charge zone has been the subject of debate in solid-state battery literature~\cite{yamamotoDynamicVisualizationElectric2010,deklerkSpaceChargeLayersAllSolidState2018}. A fundamental ingredient missing from existing space charge models is the extent of the influence of the surface on defect properties. This consideration may help resolve the debate on the detrimental aspect of the space charge zone.
While many studies have examined the differences in DFE between bulk and surface defects~\cite{freysoldtFirstprinciplesCalculationsCharged2018,ahmadUnderstandingEffectLead2022,meggiolaroFormationSurfaceDefects2019},  the complete spatial variation of the DFE and its transition from the bulk to the surface value is not well understood.

Here, we comprehensively map out the variation in  two important properties of defects from the bulk to the surface, defect formation energy (DFE) and migration energy using first-principles calculations on a prototypical solid ion conductor, \ce{Li3OCl}. \ce{Li3OCl} has been the subject of numerous studies on phase stability, ion transport mechanism, stoichiometric and aliovalent doping, and interface wettability with lithium metal~\cite{emly2013phase,lu2015defect,mouta2014concentration,stegmaierLiDefectsSolidState2017,baktash2021effect,kim2019predicting} 
We first examine the modulation in DFE of dominant defects in \ce{Li3OCl} on moving from the bulk to the surface. We find that the DFE of lithium vacancies and interstitials at surfaces is lower than the bulk by as much as 0.85 eV, leading to a drastic increase in defect densities at surfaces. To map out the function $f(x)$ defining the relation between the DFE and the distance from the surface, $x$, we perform calculations with the defect located in different layers. We find that the DFE exponentially decays on moving from the surface to the bulk value, and the length scale of the decay is dependent on the type of defect, surface termination, and composition of the layer. The DFE function has major implications on the defect density profile near the surface and space charge layers in ceramics and semiconductors~\cite{swiftModelingElectricalDouble2021}. 
The exponential nature of the function $f(x)$ uncovered in the simulations offers a promising avenue for developing more accurate models of interfacial ion conductivity by revealing the effect of grain size on average DFE. Alongside the thermodynamic aspect of defect behaviour, our work also explores defect migration results to clarify kinetic aspects of the near-surface defects and defect redistribution. We highlight the preferential aggregation of vacancy defects towards free surfaces and the comparable migration kinetics of interstitial defects between surface and bulk regions.

Our results not only provide an understanding of point defects under realistic conditions encountered in batteries but also serve as guidelines for manipulating their properties to optimize ion transport, in particular, in the bottleneck regions involving grain boundaries and interfaces.  Our results on the implications of DFE variation apply to interfacial and surface transport in  solid ion conductors in other areas besides 
solid-state batteries~\cite{krauskopfFastChargeTransfer2020} including  electrocatalysis~\cite{Wang2022defect}, fuel cells~\cite{Ramaswamy2019alkaline}, gas sensors~\cite{Bakker2002electrochemical},  and corrosion~\cite{king2014accurate}.

\section{Methods}
\subsection{First-Principles Calculations}
Density functional theory (DFT) as implemented in Quantum Espresso~\cite{giannozziQUANTUMESPRESSOModular2009,giannozziAdvancedCapabilitiesMaterials2017a} was used to study \ce{Li3OCl} anti-perovskite solid ion conductor. The Perdew-Burke-Ernzerhof
exchange-correlation functional~\cite{perdewGeneralizedGradientApproximation1996}  with ultra-soft pseudopotentials (USPP)\cite{garrityPseudopotentialsHighthroughputDFT2014} and norm-conserving pseudopotentials (NCPP)\cite{vansettenPseudoDojoTrainingGrading2018} were used. 
Dispersion correction of the type DFT-D3 proposed by \citeauthor{Grimme2010}\cite{Grimme2010} was used to account for the van der Walls dispersion forces.
For bulk calculations, a 3×3×3 supercell consisting of 135 atoms \ce{(Li81O27Cl27)} and a 4×4×4 supercell consisting of 320 atoms \ce{(Li192O64Cl64)} were modeled using periodic boundary conditions. In order to keep the extended systems' charge neutrality, compensating background charges were used in both the interstitial and vacancy computations.
The Brillouin zone was sampled using a 2×2×2 grid for bulk and a 2×2×1 grid for slabs after testing. 
Energy convergence was achieved with these k-point sampling densities to within $\sim$0.001 meV/atom for USPP and within $\sim$0.002 meV/atom for NCPP. The self-consistency loop's energy threshold for convergence was set at $10^{-4}$ Ry, while the force threshold for the relaxation of the ionic positions was set at $10^{-3}$ Ry/Bohr. Energy cutoffs of 40 Ry and 200 Ry were used for the wave function and electron density respectively. For all the slab models, a vacuum length of 10 {\AA} was used on each side after testing with total vacuum lengths of 15, 20, and 25 {\AA}. The effect of vacuum on the DFE is provided in Table S2. For all the surface DFE calculations we used a 4 by 4 slab consisting of 19 layers except for the dependence of surface termination (\autoref{fig:surf-term}) where we used a 3 by 3 slab consisting of 19 layers. For supercell and surface generation, we used the pymatgen~\cite{Ong2013pymatgen} and atomic simulation environments~\cite{HjorthLarsen2017atomic}, and for visualization purposes we used the VESTA package~\cite{Momma2011vesta}.

\subsection{Defect Formation Energies (DFE)}

We perform simulations of two types of defects: positively charged lithium interstitials and negatively charged lithium vacancies which have been proposed as dominant charge carriers in \ce{Li3OCl}. In the Kröger-Vink notation, they are denoted as $\text{Li}_{i}^{\boldsymbol{\cdot}}$ and $V_{\text{Li}}^{'}$. Our study of these defects is based on the general principle that a pair of oppositely charged defects determine the properties of a material such as the Fermi level~\cite{swiftModelingElectricalDouble2021}.  
The formation energy of a charged \ce{Li+} defect is defined as \cite{stegmaierLiDefectsSolidState2017,freysoldtFullyInitioFiniteSize2009,kim2020quick}
\begin{equation}
E_{\text{f}}(\varepsilon_F, \mu_{\text{Li}}) = E_d - E_p - n_{\text{Li}}\mu_{\text{Li}} + q(\varepsilon_{\text{VBM}} + \varepsilon_F) + \Delta_{\text{corr}},
\label{dfe_corr}
\end{equation}
where $E_d$ and $E_p$ are the total energy of the defective and pristine structures respectively. $n_{Li}$ is the number of Li atoms added or removed from the pristine structure (bulk supercell or slab) to create the defective structures. $\mu_{Li}$ is the chemical potential of Li, $q$ is the total charge of the defective structure, $\varepsilon_{\text{VBM}}$ is the energy of the pristine structure's valence band maximum, $\varepsilon_{\text{F}}$ is the Fermi level referenced to the VBM of the pristine structure. 
For all the calculations of a \ce{Li+} vacancy, the value of $n_{Li}$ and $q$ in \autoref{dfe_corr} is -1 and for all the \ce{Li+} interstitial cases, these values are +1.  $\Delta_{\text{corr}}$ is a correction term that neutralizes the effect of background charge compensation while calculating the total energy of the structures with charged defects. This correction term also accounts for charged defect-defect interactions in the periodic images of the supercell and slabs. We followed the method proposed by \citeauthor{freysoldt2014first}\cite{freysoldtFullyInitioFiniteSize2009,freysoldt2018first} to calculate the defect correction terms for bulk and surfaces.
For the correction terms of bulk supercells (3x3x3 and 4x4x4) sxdefectalign\cite{freysoldt2014first} code was used while for slabs, the sxdefectalign2d\cite{freysoldt2018first} code was used with a dielectric constant of 15 for \ce{Li3OCl}\cite{stegmaierLiDefectsSolidState2017}.

\subsection{Defect Migration Barrier Calculations}

To determine the energy barrier for the migration of charged \ce{Li+} defects along the minimum energy path near the \hkl(001) surface terminated with LiCl and in the bulk \ce{Li3OCl}, we employed the Nudged Elastic Band (NEB) method, which is available in the Quantum Espresso software package. The force convergence criteria used for ionic relaxation was less than 0.05 eV/Å except for only one case where we used 0.06 eV/Å. In every case, we employed seven images to determine the minimum energy path and the migration energy, both for migration towards the surface and towards the bulk. The near-surface NEB calculations were performed in a \ce{LiCl} terminated 2 by 2 slab \hkl(001) consisting of 15 alternating \ce{LiCl} and \ce{Li2O} layers after testing with 2x2x2 and 3x3x3 bulk supercells. Details of the test results are provided in the Supporting Information (Fig. S7). For the surface NEB calculations, a 3x3x1 k-point mesh was employed.

\section{Results}

\begin{figure}[htbp]
    \centering
    \includegraphics[width=\textwidth]{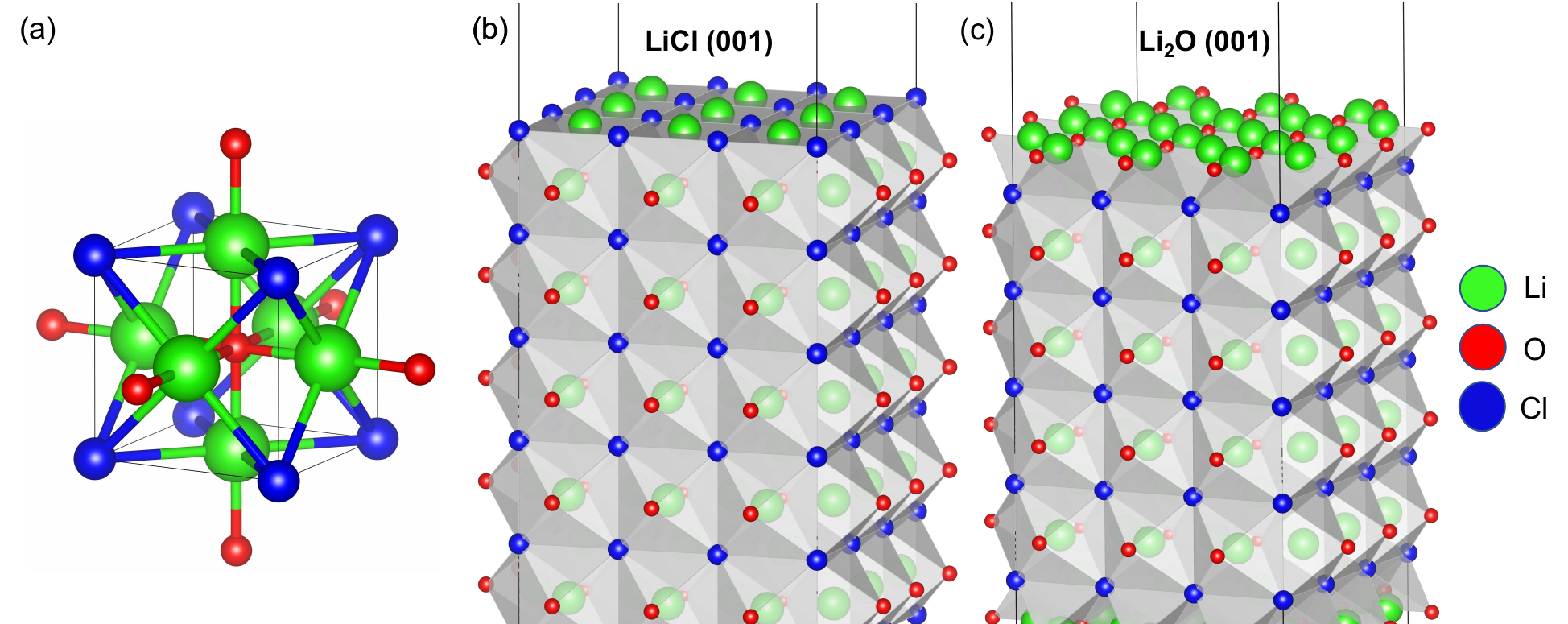}
    \caption{Structures of \ce{Li3OCl} unit cell and surfaces. (a) Primitive unit cell of \ce{Li3OCl} showing the antiperovskite structure and \hkl(001) surface of \ce{Li3OCl} with (b) LiCl termination and (c) \ce{Li2O} termination.}
    \label{fig:structure}
\end{figure}

\autoref{fig:structure} shows the structures of the primitive cell of \ce{Li3OCl} and \hkl(001) surface of \ce{Li3OCl} with  LiCl and \ce{Li2O} terminations considered in this work. \ce{Li3OCl} forms an anti-perovskite structure with the O atom surrounded by six equivalent Li atoms at face centers forming \ce{OLi6} octahedra and 8 equivalent Cl atoms on the corners of the cube. 
LiCl termination has a lower surface energy and higher work function as shown in \autoref{tab:surf} under Li-rich conditions in agreement with previous work\cite{kim2019predicting,wu2020first}. %
Additional information regarding the procedures for determining surface energy and work function can be found in the Supporting Information.

\begin{table}[htbp]
    \centering
\begin{tabular}{ |c|c|c| } 
 \hline
 Surface termination & Surface energy (J/m$^2$) & Work function (eV)\\
 \hline
LiCl & 0.327 & 2.544 \\ 
\ce{Li2O} & 0.984 & 2.361 \\ 
 \hline
\end{tabular}
    \caption{Properties of \hkl(001) \ce{Li3OCl} surfaces.}
    \label{tab:surf}
\end{table}

\subsection{Defect Configurations}

Before performing DFE calculations, it is important to identify the correct ground state structure for \ce{Li+} vacancies and interstitials which might involve bond distortions~\cite{mosquera-loisIdentifyingGroundState2023}. 
To introduce a \ce{Li+} vacancy defect in the \ce{Li3OCl}  structure, a single Li atom is removed from its original lattice site adjacent to the \ce{OLi6} octahedra.\cite{emly2013phase} 
This specific configuration, determined to have the lowest energy, was consistently employed in all our vacancy defect-related calculations.

\begin{figure}[htbp]
    \centering
    \includegraphics[width=\textwidth]{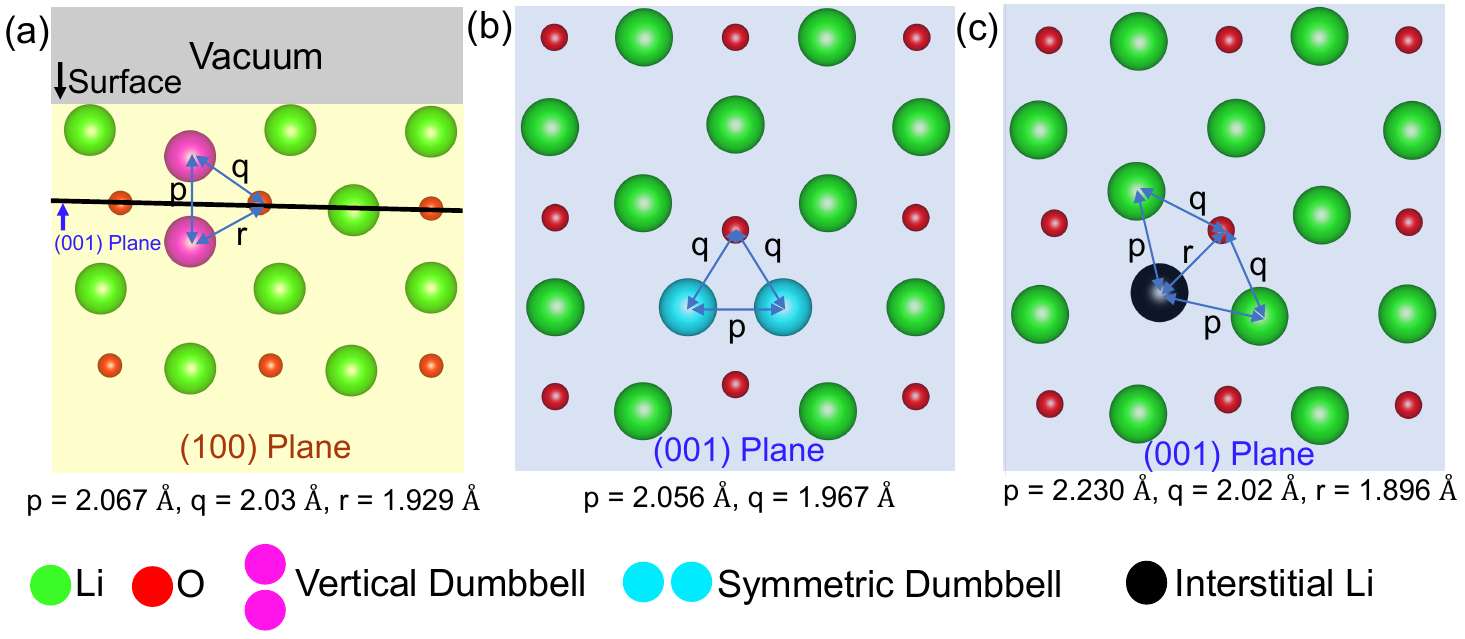}
    \caption{Ball and stick representation of three \ce{Li+} interstitial defect configurations near LiCl terminated \hkl(001) surface. Cl atoms are not shown in the figure for clarity.
    (a) Vertical dumbbell configuration perpendicular to the \hkl(001) plane (or the plane of \ce{Li2O} layer). (b) energetically favorable symmetric dumbbell configuration within the \ce{Li2O} layer. 
    (c) a higher energy configuration, featuring an interstitial Li (black) located equidistant from the two nearest Li atoms and the formation of a shorter bond length with the nearest O atom.}
    \label{int_3}
\end{figure}

The interstitial Li in bulk \ce{Li3OCl} forms a dumbbell structure with a Li atom of an \ce{OLi6} octahedron. The center of the dumbbell lies at the octahedral corner where the Li-ion has been displaced.\cite{emly2013phase}
In our investigation of interstitial defects near \ce{Li3OCl} surface, we examined three distinct configurations, illustrated in \autoref{int_3}. The dumbbell can be oriented perpendicular (\autoref{int_3}a) or parallel (\autoref{int_3}b) to the surface. A third interstitial configuration is possible where the additional Li atom is equidistant from two other Li atoms (\autoref{int_3}c). The interstitial Li in this configuration,  forms a shorter bond with the oxygen atom.
The perpendicular/vertical dumbbell configuration ceases to be symmetric due to the different distances of the Li atoms in the dumbbell from the surface. Notably, the symmetric dumbbell configuration (\autoref{int_3}b), within the \ce{Li2O} layer exhibited the lowest energy among the observed configurations. The vertical dumbbell configuration is higher in energy than the symmetric dumbbell configuration by 138 meV.  The configuration depicted in \autoref{int_3}c has the highest energy, 293 meV above the symmetric dumbbell configuration. 
For all our subsequent calculations of defects in different layers, we focused on two defect configurations: vertical and symmetric dumbbells due to their low energy. The symmetric dumbbell is referred to as the interstitial in the \ce{Li2O} layer while the vertical dumbbell is referred to as the interstitial in the LiCl layer.

\subsection{Electronic Structure}

\begin{figure}[htbp]
    \centering
    \includegraphics[width=\textwidth]{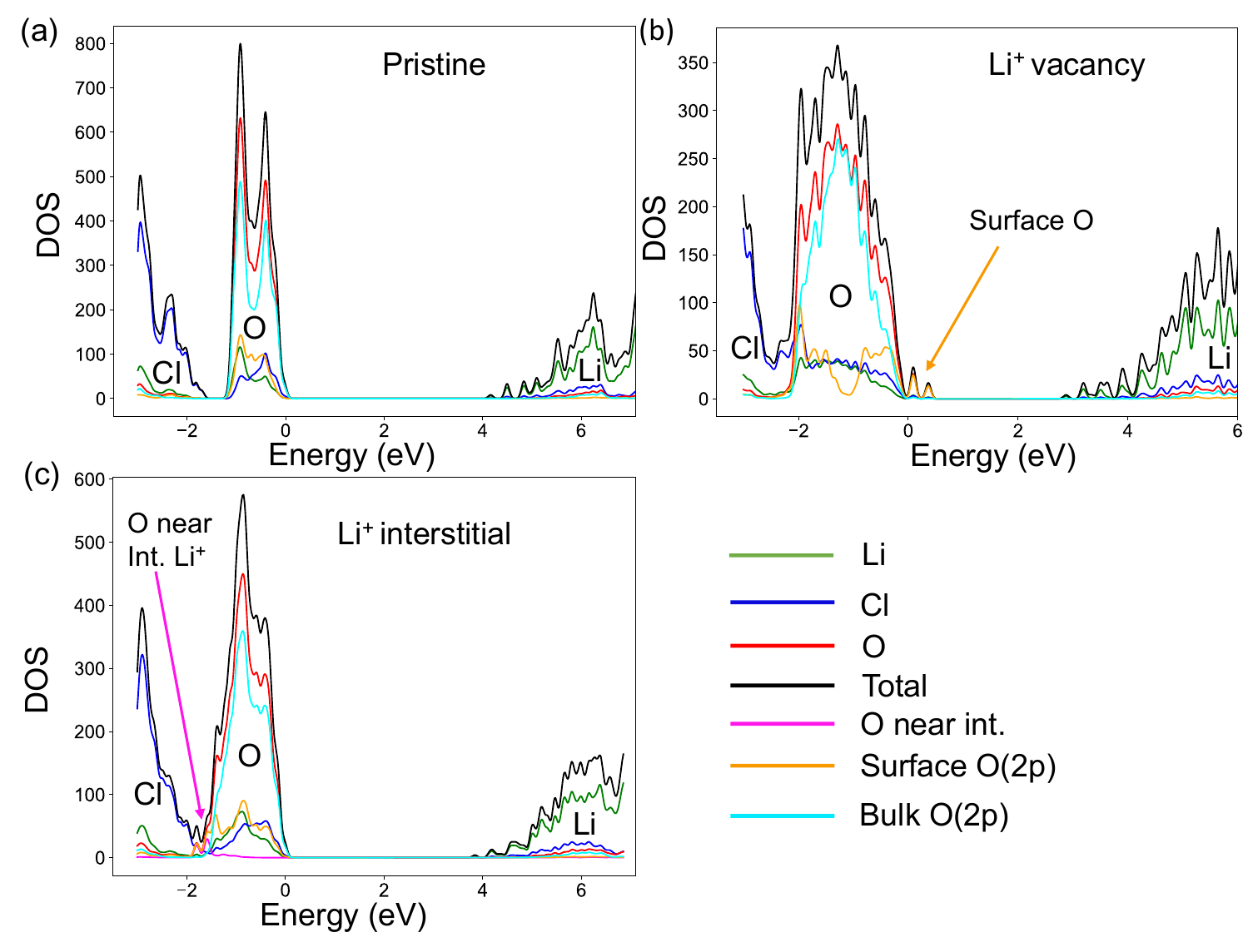}
    \caption{Total (black) and projected (colored) density of states for LiCl terminated symmetric (001) (a) pristine slab, and slabs (b) with surface \ce{Li+} vacancy and (c) surface \ce{Li+} interstitial. The labels Li, Cl, O, etc. mark the individual contributions of different atoms and orbitals to the DOS. The contribution of the oxygen is divided into bulk and surface oxygen states. The contribution of the oxygen closest to the defect is also marked with arrows in b and c. The newly generated  states due to oxygen near the \ce{Li+} vacancy defect  cause a reduction in the band gap by 0.82 eV.}
    \label{fig:dos}
\end{figure}

The differences in the electronic structure of pristine and defective \ce{Li3OCl} with \ce{Li+} vacancies and interstitials in the bulk structure have been demonstrated by \citeauthor{stegmaierLiDefectsSolidState2017}~\cite{stegmaierLiDefectsSolidState2017}.  In \autoref{fig:dos}, we plot the electronic density of states (DOS) of the LiCl terminated  (001) surface in the pristine state and compare it with the two defective states containing a \ce{Li+} vacancy and interstitial at the surface.
The valence band maximum is mainly composed of O 2p states. The conduction band minimum is composed of Li 2p states, with small contributions from Li 2s and Cl 3s states (see also Fig. S2). We differentiate the contribution of the oxygen atoms to the DOS into surface and bulk contributions. The oxygen atoms in the layer nearest to the surface are denoted as surface oxygen and those located deeper in the surface are denoted as bulk oxygen. The nature of surface and bulk oxygen contributions to the DOS changes in the defective structures. In the pristine slab, the oxygen from the bulk contributes to the VBM while in the slab with the vacancy defect, additional states are generated at the VBM due to surface oxygen. These states are marked by arrow in \autoref{fig:dos}b. This reduces the band gap by 0.82 eV. 
In comparison, an interstitial \ce{Li+} defect does not substantially affect the band gap, lowering it by only 0.26 eV (\autoref{fig:dos}c). %
The oxygen atoms bonded with the \ce{Li+} interstitial do not contribute to the band edges but to states located $\sim$2 eV lower than VBM as shown by the arrow in \autoref{fig:dos}c. This behavior is similar to the interstitial defect in the bulk~\cite{stegmaierLiDefectsSolidState2017}.

\subsection{Relationship between DFE and Distance from the Surface}
\begin{figure}[htbp]
    \centering
    \includegraphics[width=\textwidth]{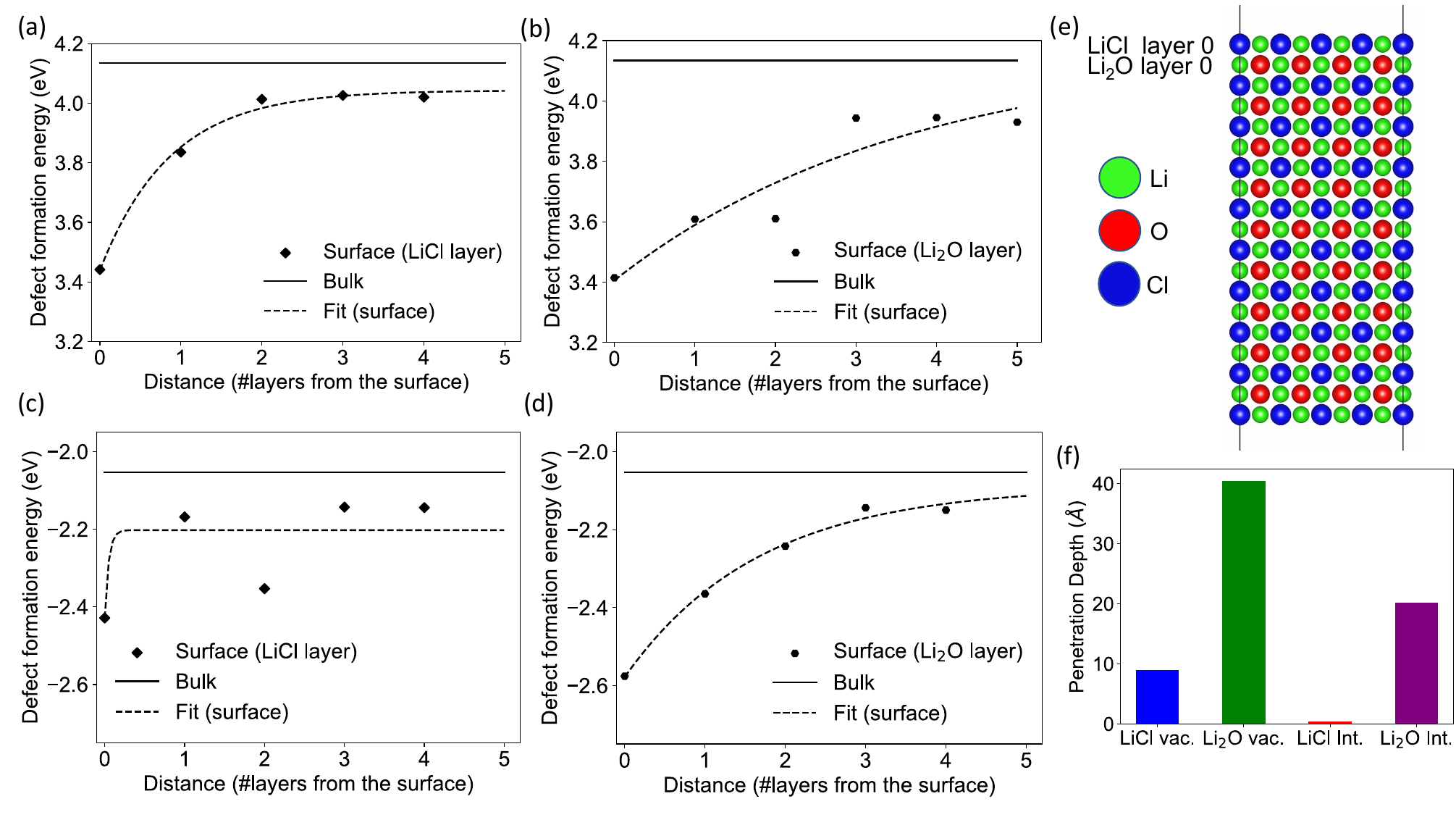}
    \caption{DFE as a function of distance (in terms of number of layers) from the surface of LiCl terminated slab. The VBM is used as the Fermi level. Vacancy DFEs in (a) LiCl layers and (b) \ce{Li2O} layers.  Interstitial DFEs in  (c) LiCl and (d) \ce{Li2O} layers.  (e) Illustration of the numbering of layers from the surface of the slab that was used for DFE calculations. (f) A bar chart showing the penetration depth at which the DFEs in the slab reach 99\% of their converged value. Defects in \ce{Li2O} layers have a higher penetration depth. The DFEs in (a-d) are calculated using a 4 by 4 slab consisting of 19 layers except for b where the surface and deepest layer calculations were performed using a thicker 23 layer slab. Similar calculations performed using a 19 layer 3  by 3  slab are shown in Fig. S3 for comparison.}
    \label{fig:LiClsurf}
\end{figure}

We first focus on the more stable LiCl termination to comprehensively investigate the dependence of DFE on the distance from the surface in the alternating LiCl and \ce{Li2O} layers. 
In \autoref{fig:LiClsurf}a and b, we plot the vacancy DFE while in \autoref{fig:LiClsurf}c and d, we plot the interstitial DFE in the LiCl and \ce{Li2O} layers separately as a function of the distance (in terms of the number of layers) away from the surface. The Fermi level is set to the VBM. 
In all cases, we find that 1) the surface DFE is lower than the bulk DFE and 2) the DFE increases with the distance from the surface and eventually reaches a converged value. The DFE data can be fitted to the exponential function,
\begin{eqnarray}\label{eq:exp}
    f(x) = A + B\exp(-\eta x),
\end{eqnarray}
where $x$ is the distance from the surface. Here $A$ is the converged value of DFE, $A+B$ is the value of surface DFE, and $1/\eta$ defines the length scale for convergence. \autoref{tab:constants} lists the values of the fitting parameters for the vacancy and interstitial defects in \ce{LiCl} and \ce{Li2O} layers.

\begin{table}[htbp]
    \centering
\begin{tabular}{ |c|c|c|c| } 
 \hline
 Defect and Layer & A (eV) & B (eV) & $1/\eta$ \ce{(nm)} \\
 \hline
Vacancy in LiCl layers & 4.04  & -0.60 & 0.332\\  
 \hline
Vacancy in \ce{Li2O} layers & 4.17 & -0.76 & 1.391 \\  
 \hline
 Interstitial in LiCl layers & -2.20 & -0.23 & 0.019 \\  
 \hline
Interstitial in \ce{Li2O} layers & -2.09 & -0.49 & 0.641\\  
 \hline
 
\end{tabular}
    \caption{{A, B and $1/\eta$ values of function $f(x)$ [\autoref{eq:exp}] for vacancy and interstitial defects in LiCl and \ce{Li2O} layers.}}
    \label{tab:constants}
\end{table}

The fit quantifies the differences between the types of defects and layers in their behavior near the surface. This exponential decay of the DFE motivates the idea of a \textit{penetration depth}, which we define as the distance at which the DFE reaches 99\% of the converged value.  The penetration depth indicates the distance within the material over which the surface effects on defects persist inside the ion conductor. 
 \autoref{fig:LiClsurf}f shows the value of this penetration depth for defects in LiCl and \ce{Li2O} layers. We obtain a penetration depth of  8.97 {\AA} and 0.44 {\AA} for vacancies and interstitial defects respectively in the LiCl layers. For the \ce{Li2O} layers, the penetration depth is higher, 40.41 {\AA} for the vacancy and 20.22 {\AA} for the interstitial respectively, indicating a greater influence of the surface.

To gain insights into the factors responsible for stabilizing surface defects, we performed DFT calculations of surface defects on the  LiCl termination without structural relaxation.
This trend obtained for DFE without structural relaxation is plotted in Fig. S6(a). We find that the surface DFE is slightly higher than the DFE in the innermost layer. Beyond the surface layer, the DFE monotonically decreases with distance from the surface. This trend is opposite to the trend obtained when performing structural relaxation as shown in \autoref{fig:LiClsurf}. This indicates that significant structural relaxation occurring near surface defects is responsible for stabilizing them.

We find consistently that there is a difference of $\sim 0.2$ eV between the DFE of the deepest layer in the slab and the bulk DFE. We verified that this difference persists even when increasing the number of layers in the slab.  The discrepancy may be due to different structural relaxation of the slab compared to the bulk.
An incomplete charged defect correction at surfaces might also contribute to the difference~\cite{freysoldt2018first}.

\subsection{Dependence on Surface Termination}

\begin{figure}
    \centering
    \includegraphics[width=\textwidth]{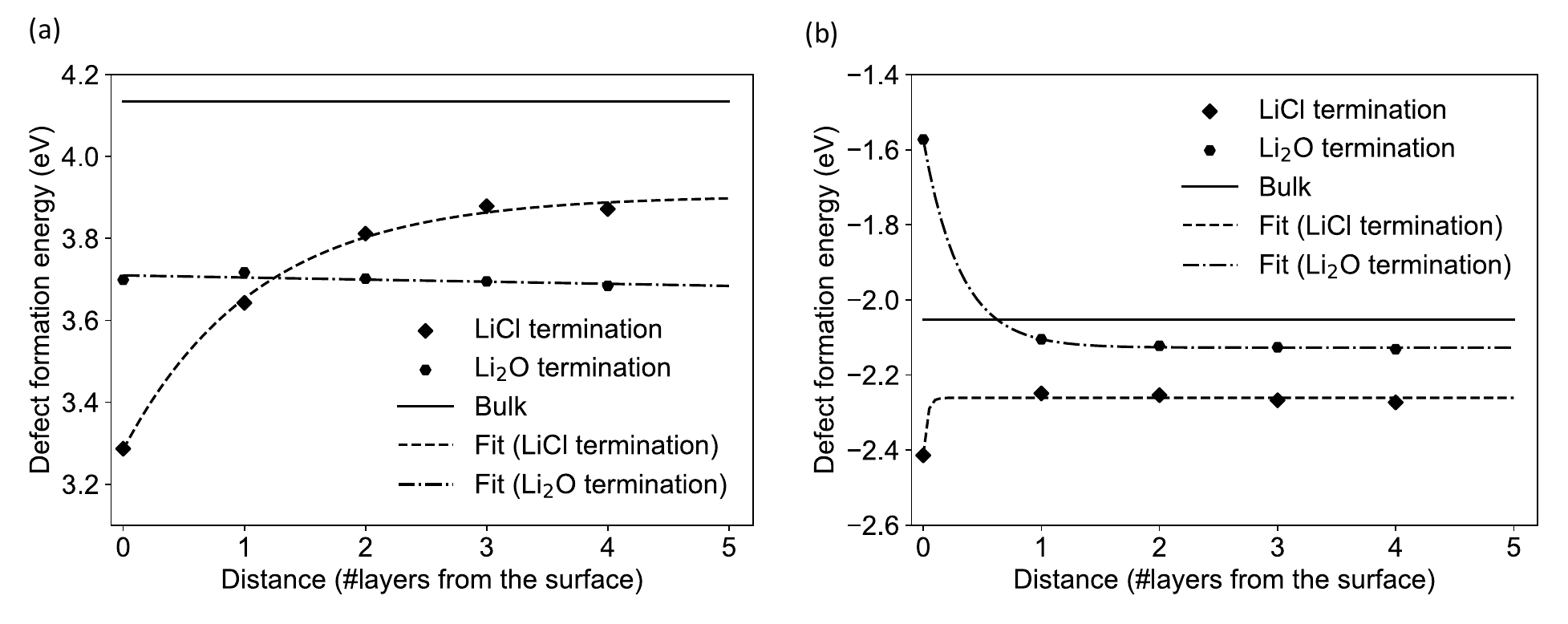}
    \caption{Comparison of defect formation energies (DFEs) for the (a) vacancy and (b) interstitial \ce{Li+} defects at LiCl and \ce{Li2O} terminated \hkl(001) surfaces of \ce{Li3OCl}. The DFE is plotted as a function of the distance from the surface in terms of the number of layers.  The surface DFE is generally lower than the bulk DFE except for \ce{Li+} interstitial defect at \ce{Li2O} terminated surface.}
    \label{fig:surf-term}
\end{figure}

Next, we compare the DFE for LiCl and \ce{Li2O} terminated surfaces as a function of the distance from the surface. 
\autoref{fig:surf-term}a and b plot the variation in the DFE as a function of the distance from the surface for the vacancy and interstitial defects respectively. 
 For the vacancy defect, the surface DFE is lower than the bulk for both the LiCl and \ce{Li2O} terminations. The difference is as high as 0.85 eV for the LiCl termination. This suggests that, thermodynamically, \ce{Li+} vacancies will aggregate towards free surfaces such as grain boundaries away from the bulk. The concentration of vacancies at surfaces can be $ \exp(\Delta E/k_B T)\approx 10^{14}$ times that in the bulk ($\Delta E = 0.85$ eV, $k_BT= 26$ meV at 300 K). Defect-defect interactions may become prominent at such high concentrations.
While the vacancies in the \ce{LiCl} terminated slab show an exponential variation in the DFE from the surface to the inner bulk-like layers, the DFE for the vacancies in the \ce{Li2O} terminated slab remains nearly constant with layer distance. 

Similar to the vacancies, we find that the interstitial DFE at the surface differs from the bulk DFE. The LiCl termination has a lower surface DFE while the \ce{Li2O} termination has a higher surface DFE than the bulk, in contrast to all other cases.  As a result, the \ce{Li2O} surface will have a lower concentration of interstitial defects compared to the bulk, making vacancies the primary charge carriers.

\subsection{Structural Relaxation near Defects: Bulk vs. Surface}
\begin{figure}[htbp]
    \centering
    \includegraphics[width=\textwidth]{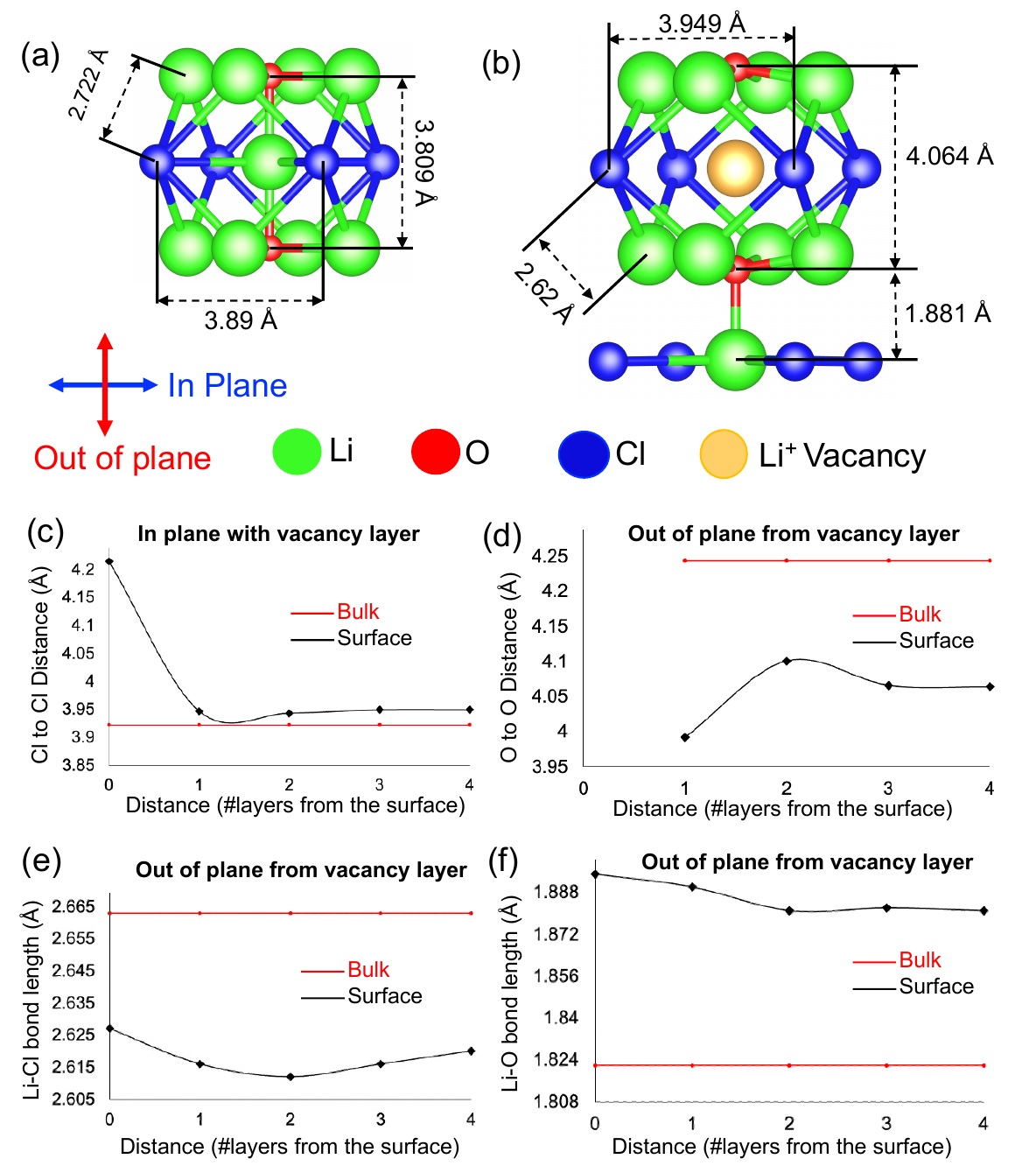}
    \caption{
   (a) The in plane Cl to Cl distance, out of plane O to O distance and Li-Cl bond length in a pristine slab, (b) effect of geometric relaxation due to \ce{Li+} vacancy in the deepest LiCl layer on the distances shown in (a) and out of plane Li-O bond length. Anions (Cl, O) move away and cations (Li) move in towards the \ce{Li+} vacancy site. Structural distortion effects due to the geometric relaxation on (c) in plane Cl to Cl distance, (d) out of plane O to O distance, (e) out of plane Li-Cl bond length from the LiCl layer containing vacancy and (f) out of plane Li-O bond length in between the next \ce{Li2O} layer and LiCl layer as a function of layer number from the surface of a LiCl terminated 4 by 4 slab. }
    \label{fig:distance}
\end{figure}

To explain the differences in surface DFEs across the layers and with the bulk, we performed an analysis of the local bonding environment around the vacancy defect in the LiCl layers at the LiCl-terminated surface. When a \ce{Li+} vacancy is generated, the neighboring O and Cl atoms move away from the vacancy while the neighboring Li atoms move towards the vacancy. The Cl atoms are located in the plane of the layer containing the defect while the O and Li atoms are located out of the plane as shown in \autoref{fig:distance}(a-b).     We plot the variation in the in plane Cl-Cl distance, out of plane O-O distance, out of plane Li-Cl and Li-O bond lengths in \autoref{fig:distance}(c-f) in the different layers together with the respective values for the defect in the bulk. %
The variation occurs due to different degrees of structural relaxation in the layers.
Based on the relative comparison of these distances and bond lengths as a function of layer number (Fig. S5), we observed that the changes in the in plane Cl to Cl distances appear dominant near the surface.
The Cl-Cl distance is the highest in the outermost layer and then becomes nearly constant. 
The O-O distance exhibits an opposite trend; these distances increase from the surface on moving towards the inner layers. Li-Cl bond lengths undergo a slight decrease from the surface towards the inner layers but then increase again near the innermost layer. The Li-O bond lengths close to the defect decrease on moving towards the inner layers and eventually reach a constant value.   All the bond lengths and distances eventually converge, mirroring the trend in DFE.

These local structural distortions lead to different bonding environments near the defect contributing to the exponential trend in vacancy DFE in the LiCl layers. We believe the same reasoning applies to other cases presented in \autoref{fig:LiClsurf}.
In addition,  the converged values of bond distances differ slightly from the bulk values. The differences may be due to an additional freedom of out of plane relaxation in slabs which does not exist in the bulk where the lattice parameters are fixed during relaxation. This is supported by the fact that the innermost out of plane O-O distance, Li-O and Li-Cl bond lengths show a larger deviation from the bulk value compared to the innermost in plane Cl-Cl distance.
These differences contribute to the deviation in DFE of the innermost surface layer from the bulk value.

\subsection{Implications for Solid ion conductors}
The differences in the DFEs between the bulk and the surface lead to the redistribution of defects within the solid ion conductor. The defects accumulate near surfaces/grain boundary core due to lower DFE and get depleted away from the surface. This is the origin of the space charge effect, which controls the ionic conductivity of many solids~\cite{maierDefectChemistryIonic1987}. Current models for space charge do not incorporate the spatial extent of the core region and treat it as completely localized at the surface~\cite{gobelNumericalCalculationsSpace2014, mebaneGeneralisedSpacechargeTheory2015,chenElectrochemomechanicalChargeCarrier2021}.
The penetration depth we calculated for surface defects can be interpreted as the spatial extent of the core region, i.e., the region where the defects accumulate near the surface.  We predict a new exponential relationship for the DFE that can be used to calculate defect density variation in the core region. The surface defect accumulation and density variation in the core region will modify the interfacial kinetics of ion transfer during battery cycling.
In addition, our result emphasizes the need for discrete space charge modeling due to the explicit inclusion of layer number dependence of DFE~\cite{xiaoDiscreteModelingIonic2022,armstrongDoubleLayerStructure1997}. The changes in defect distribution in the core and space charge zone can be modulated by grain size and processing techniques and significantly influence ionic conductivity.

Macroscopic properties, for example, ionic conductivity and activation energy, depend on the behavior of defects in all layers within the material. Based on this assumption, \citealt{meggiolaroFormationSurfaceDefects2019} developed  a model for the average DFE by weighing the contributions of two types of defects, surface and bulk,

\begin{equation}
{\text{DFE}_{\text{av}}}= (g_{\text{surf}} \cdot {\text{DFE}}_{\text{surf}} + g_{\text{bulk}} \cdot {\text{DFE}}_{\text{bulk}})/(g_{\text{surf}} + g_{\text{bulk}}),
\label{eq:angelis}
\end{equation}

where $g_{\text{surf}}$ ($g_{\text{bulk}}$) and $\text{DFE}_{\text{surf}}$ ($\text{DFE}_{\text{bulk}}$) are the number density and formation energies of defects at the surface (bulk). In their study, all defects except for surface defects were categorized as bulk defects.

\begin{figure}[htbp]
    \centering
    \includegraphics[width=\textwidth]{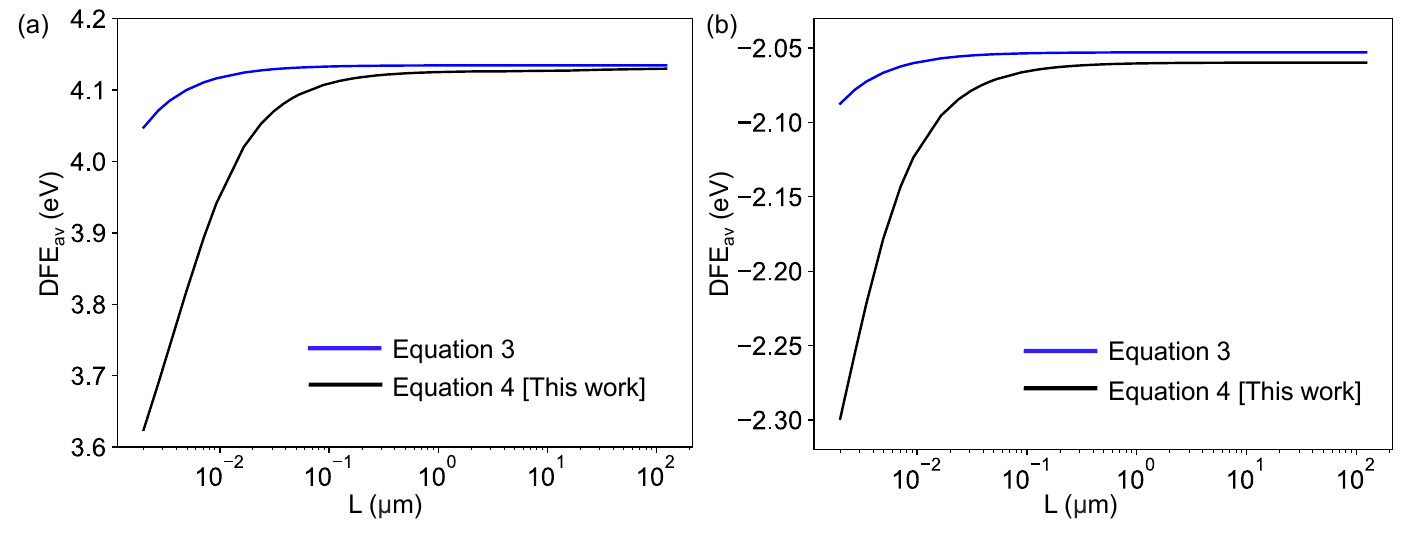}
    \caption{Comparison of average DFE between the model proposed by \citealt{meggiolaro2019formation} (blue) and this work which calculates DFE as a function of layer distance (black). Average DFE for (a) \ce{Li+} vacancy and (b) \ce{Li+} interstitial as a function of grain thickness starting from 2 nm (11 layers) to 122 {$\mu$}m (672,364 layers).}
    \label{fig:avg_dfe}
\end{figure}

Our results point to the presence of more than two types of defects with properties depending on the layer distance from the surface. Using the fitted equations from \autoref{fig:LiClsurf}, we revised the average DFE model by incorporating the exponential trend of DFE in different layers,

\begin{equation}
\text{DFE}_{\text{av}}= \frac{\sum_{i=0}^{n} g^{\ce{LiCl}}(i) \cdot \text{DFE}^{\ce{LiCl}}(i) + \sum_{i=0}^{n-1}g^{\ce{Li2O}}(i) \cdot \text{DFE}^{\ce{Li2O}}(i) } {\sum_{i=0}^{n} g^{\ce{LiCl}}(i) + \sum_{i=0}^{n-1} g^{\ce{Li2O}}(i)}
\label{eq:our_exp}
\end{equation}

where $g^{\ce{LiCl}}(i)$ and $g^{\ce{Li2O}}(i)$ 
are the density of defects in the $i^{th}$ \ce{LiCl} and \ce{Li2O} layer respectively. $\text{DFE}^{\ce{LiCl}}(i)$ and $\text{DFE}^{\ce{Li2O}}(i)$ are the DFEs in the $i^{th}$ \ce{LiCl} and \ce{Li2O} layer respectively. $2n+1$ is the total number of layers. We restrict ourselves to a quasi-1D model with the layers forming planes perpendicular to the surface, sufficient to get insights into defect behavior~\cite{xiaoDiscreteModelingIonic2022}. The number of layers and hence the average DFE can be controlled through the grain size of the solid ion conductor. Smaller grains contain a higher proportion of surface-like defects while larger grains contain more bulk-like defects. \autoref{fig:avg_dfe}  compares the average DFE obtained using \autoref{eq:angelis} and our revised model, \autoref{eq:our_exp}. For grain sizes smaller than 1 ${\mu}m$, surface effects dominate for both vacancy and interstitial defects.
Our model predicts a lower average DFE compared to \citealt{meggiolaroFormationSurfaceDefects2019} for smaller grain sizes. 
Based on the average DFE, the error in defect density $\sim \exp(-\text{DFE}_{\text{av}}/k_B T)$ that arises by ignoring the variation of DFE with distance exceeds 50\% for grain sizes $\leq$ 1.26 $\mu$m for vacancy defects and $\leq$ 0.46 $\mu$m for interstitial defects.

\subsection{Defect Migration}

While DFE has a significant impact on the ionic transport in solid ion conductors by controlling the defect distribution, another important property that influences it is defect mobility. Here, we quantify it using the defect migration/hopping barrier between neighboring sites in the bulk and near the surface. The migration barrier also controls the kinetics of defect redistribution predicted by us as new surfaces are generated. A difference in the migration barrier away or towards the surface is expected due to the broken symmetry at surfaces. 

\begin{figure}[htbp]
    \centering
    \includegraphics[width=\textwidth]{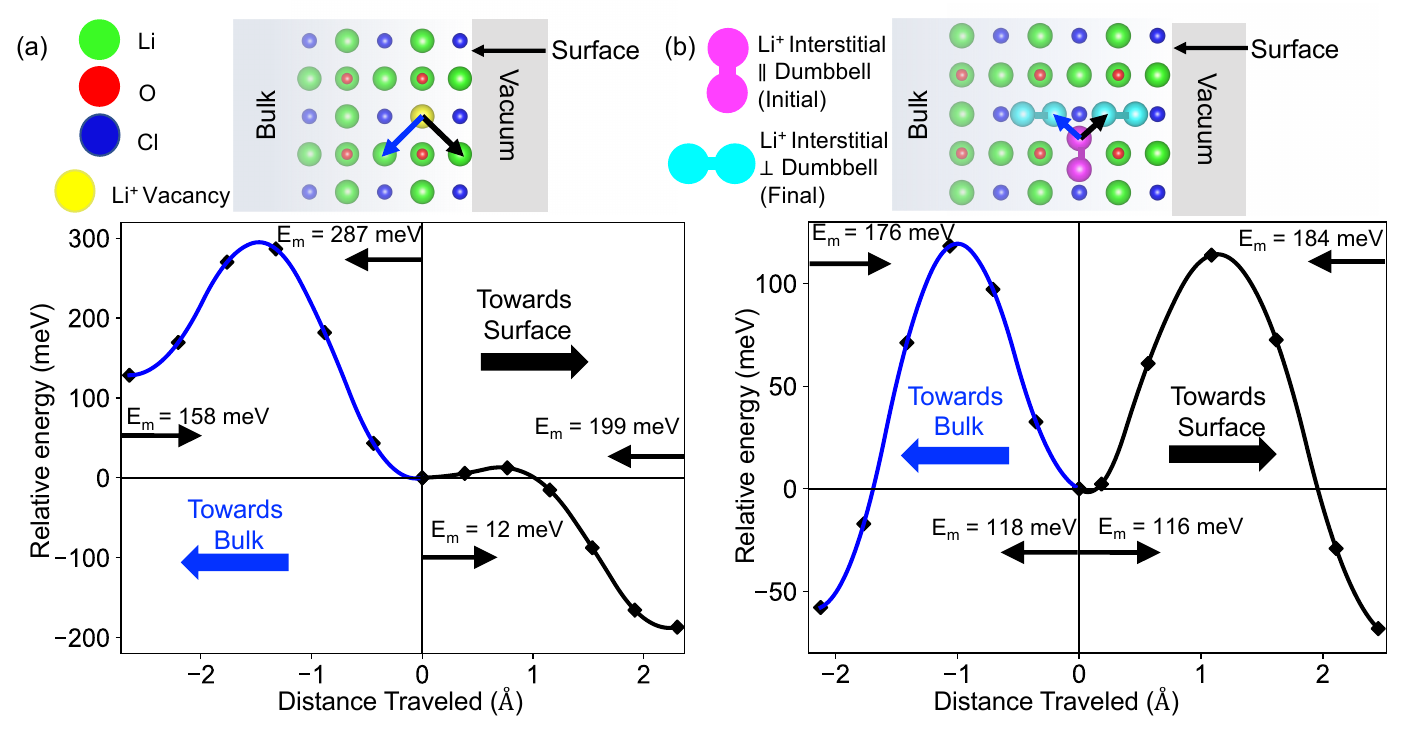}
    \caption{Comparison of (a) \ce{Li+} vacancy and (b) \ce{Li+} interstitial defect's migration energy {$E_m$} `towards surface' vs. `towards bulk'. The considered migration plane of the defects are perpendicular to the surface. For vacancy migration from a subsurface \ce{Li2O} layer {$E_\text{m}$} = 12 meV towards the LiCl terminated surface and {$E_\text{m}$} = 287 meV towards bulk-like inner LiCl layer.  %
    Interstitial defect migration is less sensitive to the surface properties since both `towards surface' and `towards bulk' cases have almost the same migration energy.
}
    \label{fig:migration}
\end{figure}

In \ce{Li3OCl}, the negatively charged Li vacancies migrate along the edge of an oxygen octahedron.  From our DFT calculations, we obtain a barrier for this migration of 302 meV in bulk \ce{Li3OCl} in agreement with previous work by \citealt{emly2013phase} that predicted a barrier of 310 meV.
To determine the kinetics of defect redistribution between the bulk and surfaces, we compared the energies for vacancy migration towards the surface and towards the bulk from a layer near the surface (2nd layer from the surface). The energy landscape for this migration is highly asymmetric and is shown in \autoref{fig:migration}a.
The migration energy is 287 meV towards the bulk and 12 meV towards the surface layer.  Both barriers are lower than the bulk value of 302 meV. The low migration barriers suggest fast kinetics of lithium redistribution towards the surface.

The migration of \ce{Li+} interstitials proceeds along a minimum energy pathway involving a 3-atom movement, the swapping of a Li atom in a lattice site with a Li dumbbell.\cite{emly2013phase, mouta2014concentration} We obtain a  barrier of 151 meV for this migration in the bulk from our DFT calculations which is close to the reported value of 145 meV.\cite{emly2013phase} 
Next, we proceeded to calculate the migration barrier near the surface layers. We found that the interstitial with dumbbell configuration is unstable on the surface layer, therefore, we calculated the migration barriers for "towards surface" and "towards bulk" pathways using the interstitial located in the next three layers away from the surface.  
For the migration of the interstitial dumbbell from its initial to final configurations shown in \autoref{fig:migration}b, the energy barrier is 116 meV towards the surface, compared to 118 meV towards the bulk. This similarity in activation energies for interstitial migration suggests that the mechanisms governing interstitial diffusion are less influenced by surface effects compared to vacancy migration. Additionally, both types of defect migration near the surface exhibit asymmetric energy landscapes, contrasting with the symmetric nature of these curves in the bulk \ce{Li3OCl}. This asymmetry may arise due to the site energetics and structural arrangements near the surface, leading to preferential migration pathways and barriers that differ from those observed in the bulk material. Our results also demonstrate the kinetics of defect redistribution are more or less spatially uniform for interstitials whereas a high degree of variability is present for vacancies, depending on the defect site and direction of movement. For example, \ce{Li+} vacancies in the 2nd layer can quickly accumulate at the surface once it is created.

\section{Summary and Conclusions}
In conclusion, we have bridged the gap between surface and bulk defects in solid ion conductors by comprehensively mapping the DFEs as a function of the distance from the surface.  
The distinct nature of these near-surface defects is expected to play an important role in the carrier distribution and transport at surfaces and interfaces in functional materials within devices.
We propose that for a comprehensive assessment of a solid ion conductor in a device, it is essential to understand the complete DFE function, $f(x)$ rather than focusing solely on bulk conductivity.
The intrinsic exponentially decaying nature of the DFE function revealed by our simulations paves the way for more accurate multiscale models of interfacial ion conductivity and impedance. Interlayers and grain-size engineering are possible routes to enhance conductivity by precise modification of the DFE function $f(x)$.  In addition, our DFE function can be used to obtain a more accurate description of the defect density profile in the space charge zones present near grain boundaries and interfaces.
By incorporating the DFE function into an average DFE model, we propose a revised model that can be used to obtain the effects of grain size on conductivity.

Our investigation into DFEs and defect migration barriers sheds light on both thermodynamic and kinetic aspects of defect behavior in \ce{Li3OCl} solid electrolyte. Thermodynamically, vacancy defects exhibit lower formation energies on the surface compared to the bulk, indicating a higher tendency for vacancy aggregation towards free surfaces. We find low migration barriers for vacancies to move towards the surface compared to the bulk, suggesting fast kinetics of defect redistribution. Similarly, interstitial defects also exhibit lower formation energies on the surface, indicating a preference for surface accumulation. However, in contrast to vacancies, the migration barriers for interstitial defects are comparable between the surface and bulk, implying similar kinetics for interstitial movement towards both regions. Our results provide a promising route for the design of high-performance solid ion conductors by simultaneous optimization of both thermodynamic and kinetic properties of defects.

\begin{acknowledgement}
We acknowledge Texas Tech University Mechanical Engineering Department startup grant for support of this research. We acknowledge the High Performance Computing Center (HPCC) at Texas Tech University and the Lonestar6 research allocation (DMR23017) at the Texas Advanced Computing Center (TACC) for providing computational resources that have contributed to the research results reported within this paper.

\end{acknowledgement}
\begin{suppinfo}

Details of computational methods, structural information, and additional analysis conducted.

\end{suppinfo}

\bibliography{zotero_refs,refs}


\section*{Supplementary material}

# Supporting Information:

# Heterogeneity in Point Defect Distribution and Mobility in Solid Ion Conductors

Md Salman Rabbi Limon and Zeeshan Ahmad[*]

*Department of Mechanical Engineering, Texas Tech University, Lubbock, Texas 79409, USA*

E-mail: zeeahmad@ttu.edu

## Surface Energy and Work Function Calculations:

To calculate the surface energy of the LiCl terminated and $Li_2O$ terminated 001 symmetric slabs at 0K we used the procedure described in the works of Tian et al.[1] and Thompson et al.[2]

$$\gamma = \frac{1}{2A}(E_{slab} - n_{\text{unit}} \cdot E_{\text{unit}} + \sum_i n_i \cdot \mu_i) \qquad (1)$$

Here, A is the vacuum facing surface area of one side of the slab. $E_{\text{slab}}$ is the total energy of the slab, $n_{\text{unit}}$ is the number of $Li_3OCl$ unit cell present in the slab, $E_{\text{unit}}$ is the total energy of a single unit cell, $n_i$ is the number of atoms of type i present in the slab in excess of the stoichiometric amount, and $\mu_i$ is the chemical potential of the element i. According to previous studies of Zhang et al.[3] and Emly et al.[4] $Li_3OCl$ is metastable at 0K and is susceptible to decomposition into LiCl and $Li_2O$. Our calculations indicate that decomposition of $Li_3OCl$ into LiCl and $Li_2O$ at 0K is exothermic and yields only 0.02eV/atom. For determining the chemical potentials of Cl and O we assumed equilibrium of LiCl and $Li_2O$

with Li rich condition (i.e BCC Li metal).

For the work function calculations of the LiCl and $Li_2O$ terminated symmetric surfaces the average of the electrostatic potential in the vacuum and Charge Neutral Fermi Energy of bulk $Li_3OCl$ were used. Using our bulk vacancy and interstitial DFEs as a function of Fermi level we get a Charge Neutral Fermi Level (CNFL) of 3.094 eV where the VBM is set to zero. Including the VBM of the 3x3x3 bulk supercell the CNFL yields 6.134 eV.

## Charge Correction Term Based on Supercell Size

For the $Li^+$ vacancy defect in a 3x3x3 supercell, the contribution of $\Delta_{corr}$ term in DFE is ≈0.16 eV for both the cases of ultra-soft and norm-conserving pseudo-potentials. With the reduction of defect density in the 4x4x4 supercells the $\Delta_{corr}$ contribution reduces to ≈0.1 eV. The correction term's contribution further decreases to ≈0.06 eV with a supercell size of 6x6x6. The defect formation energies ($Li^+$ vacancy) in 3x3x3, 4x4x4 and 6x6x6 supercells along with the contribution of $\Delta_{corr}$ terms in DFE are shown in Table S1. For the 3 by 3 and 4 by 4 supercell slabs (001) the contribution of $\Delta_{corr}$ terms in DFE at different layers is shown in Table S4.

Table S1: Dependence of bulk DFE and $\Delta_{corr}$ on supercell size.

| Supercell | DFE (eV) | $\Delta_{corr}$ (eV) | Band Gap (eV) |
|---|---|---|---|
| | $Li^+$ Vacancy | | |
| 3x3x3 | 4.135 | 0.162 | 4.5183 |
| 4x4x4 | 4.133 | 0.109 | 4.4925 |
| 6x6x6 | 4.101 | 0.058 | 4.4472 |

## Vacuum Region Test:

To determine the effect of total vacuum region on Defect Formation Energy (DFE), we calculated the $Li^+$ vacancy DFE on the surface of a symmetric LiCl terminated 3 by 3 slab consisting of total 11 alternating layers of LiCl and $Li_2O$. The VBM was set to Fermi level

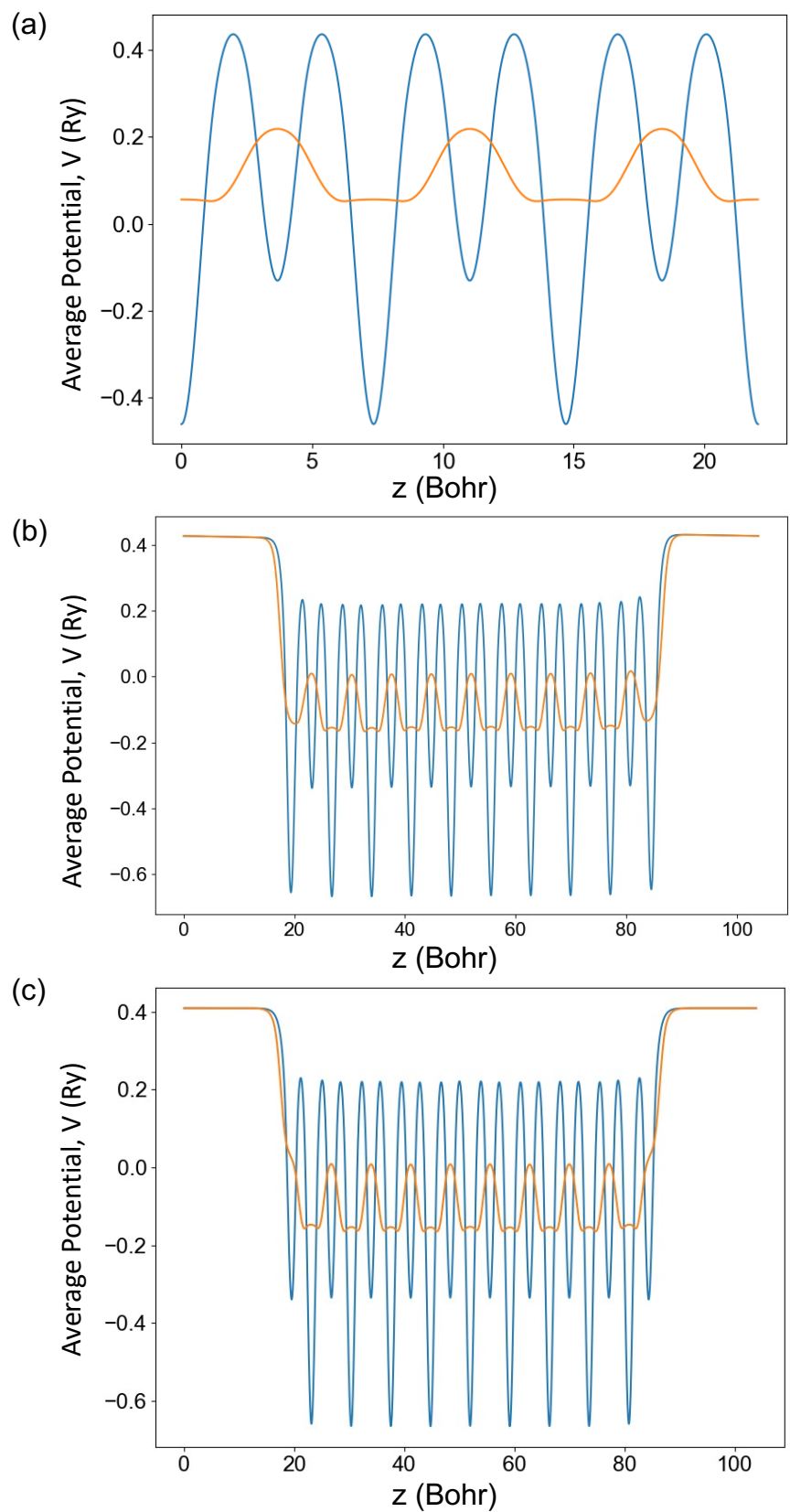

Figure S1: Average electrostatic potential [Macroscopic (orange) and Microscopic (blue)] of (a) bulk 3 by 3 supercell, (b) LiCl terminated slab, and (c) Li$_2$O terminated slab along z.

for DFE calculations. For calculating the correction terms, isolated regions used in the sxdefectalign2d[5]code were adjusted proportionally for all three total vacuum regions.

Table S2: Effect of total vacuum on DFE.

| Total Vacuum (Å) | DFE (eV) on surface | Difference (eV) with 20 Å |
|---|---|---|
| Symmetric LiCl terminated slab | | |
| 15 | 3.31 | 0.03 |
| 20 | 3.28 | 0.00 |
| 25 | 3.25 | 0.03 |

Due to negligible changes in the DFE, we choose 20 Å total vacuum for all the surface calculations.

Table S3: Comparison of $Li^+$ vacancy DFEs at different layers of LiCl and $Li_2O$ terminated symmetric 001 slabs using ultra-soft pseudo-potentials (uspp) and norm-conserving pseudo-potentials (ncpp). (3x3x3 supercell slabs)

| Layer number | DFE (eV) [uspp] | DFE (eV) [ncpp] |
|---|---|---|
| LiCl terminated | | |
| LiCl - 0 (surface) | 3.28662 | 3.35714 |
| LiCl - 1 | 3.64274 | 3.62337 |
| LiCl - 2 | 3.81190 | 3.80724 |
| LiCl - 3 | 3.87891 | 3.84677 |
| LiCl - 4 | 3.87187 | 3.83009 |
| $Li_2O$ terminated | | |
| $Li_2O$ - 0 (surface) | 3.69898 | 3.79734 |
| $Li_2O$ - 1 | 3.71724 | 3.83020 |
| $Li_2O$ - 2 | 3.70197 | 3.81808 |
| $Li_2O$ - 3 | 3.69513 | 3.78711 |
| $Li_2O$ - 4 | 3.68440 | 3.77443 |

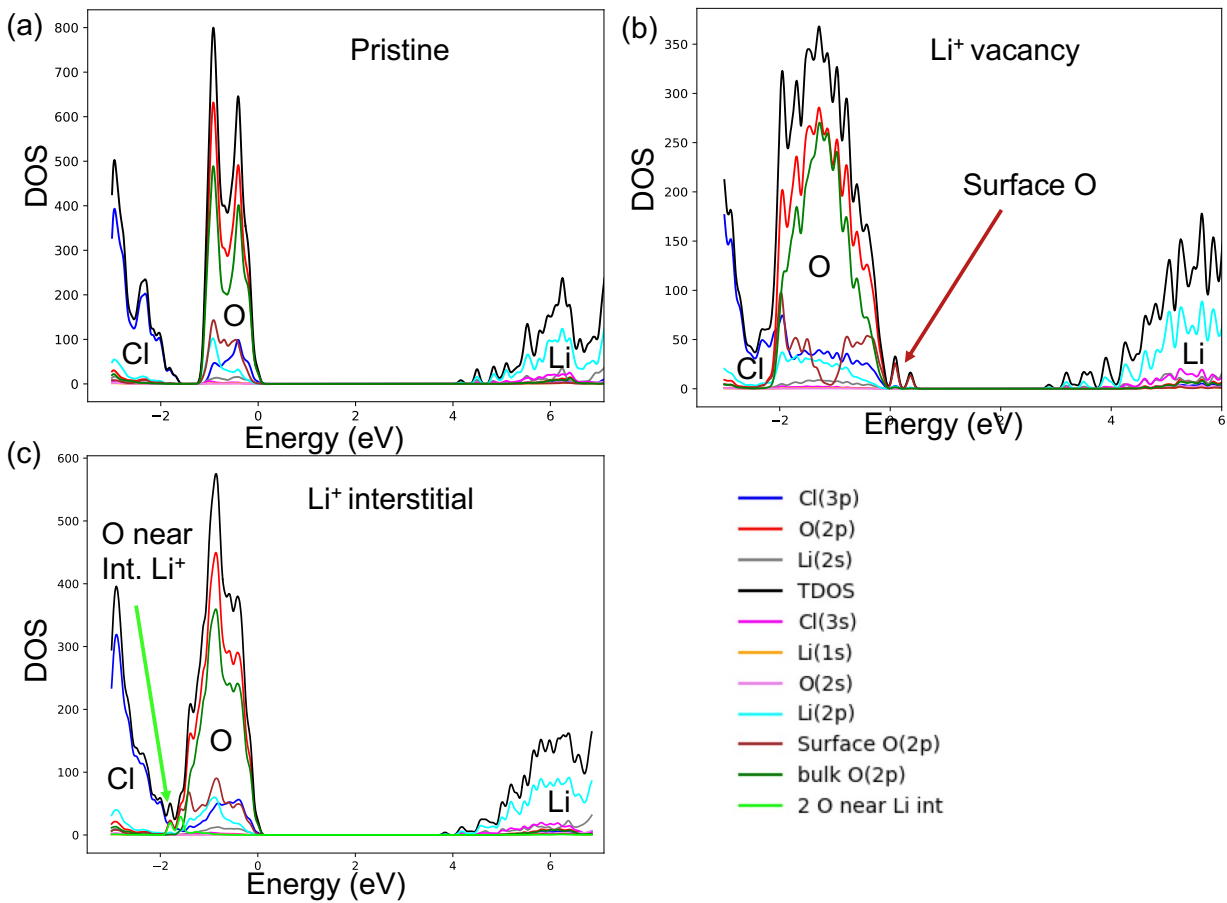

Figure S2: A more detailed version of Figure 2.

Table S4: Comparison of the $\Delta_{\text{corr}}$ term's contribution in DFE between 3x3x3 and 4x4x4 supercell slabs with LiCl termination.

| Layer number | $\Delta_{\text{corr}}$ (eV) [3x3x3] | $\Delta_{\text{corr}}$ (eV) [4x4x4] |
|---|---|---|
| Vacancy in LiCl layers | | |
| LiCl - 0 (surface) | -0.57233 | -0.27344 |
| LiCl - 1 | -0.53208 | -0.24626 |
| LiCl - 2 | -0.45766 | -0.21033 |
| LiCl - 3 | -0.41446 | -0.18510 |
| LiCl - 4 | -0.39791 | -0.17600 |
| Vacancy in Li$_2$O layers | | |
| Li$_2$O - 0 | -0.56170 | -0.28406 |
| Li$_2$O - 1 | -0.49576 | -0.24617 |
| Li$_2$O - 2 | -0.43466 | -0.22364 |
| Li$_2$O - 3 | -0.39771 | -0.17973 |
| Li$_2$O - 4 (mid-slab) | -0.38376 | -0.17280 |
| Interstitial in LiCl layers | | |
| LiCl - 0 (surface) | -0.45877 | -0.21878 |
| LiCl - 1 | -0.46859 | -0.21676 |
| LiCl - 2 | -0.43770 | -0.16842 |
| LiCl - 3 | -0.41100 | -0.19808 |
| LiCl - 4 | -0.40093 | -0.18740 |

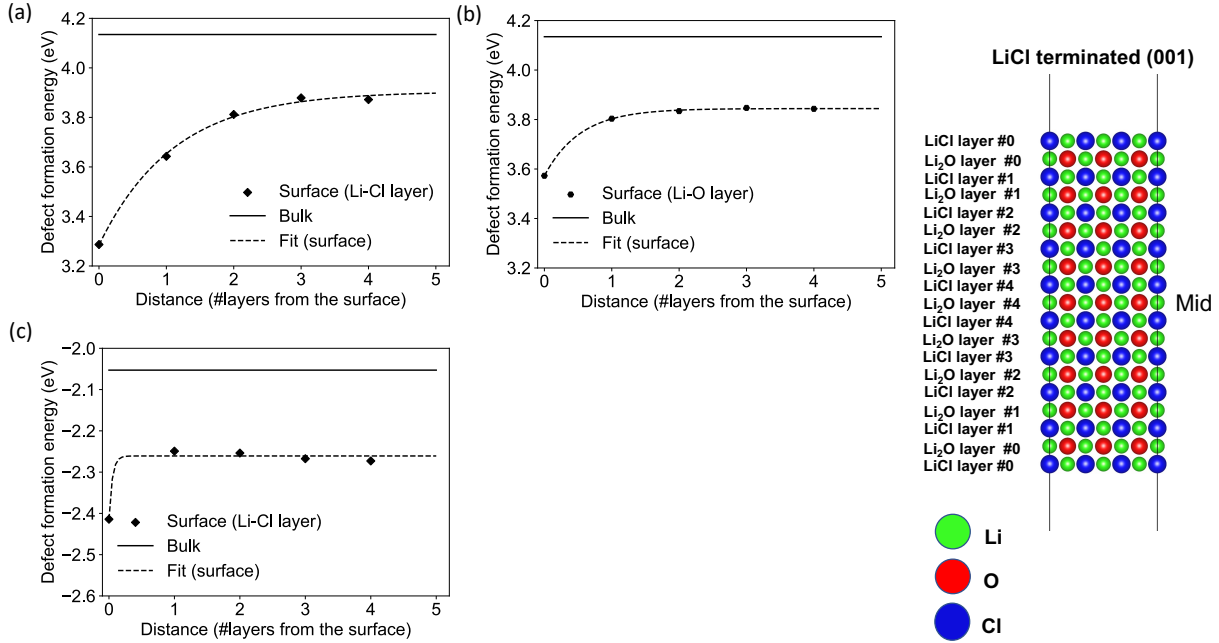

Figure S3: DFE as a function of distance (number of layers) from the surface of LiCl terminated surface. Vacancy DFEs in (a) LiCl layers and (b) Li$_2$O layers. Interstitial DFEs in (c) LiCl layers. An illustration showing the layer numbers starting from the surface of the 3 by 3 slab that was used for the DFE calculations in (a) - (c).

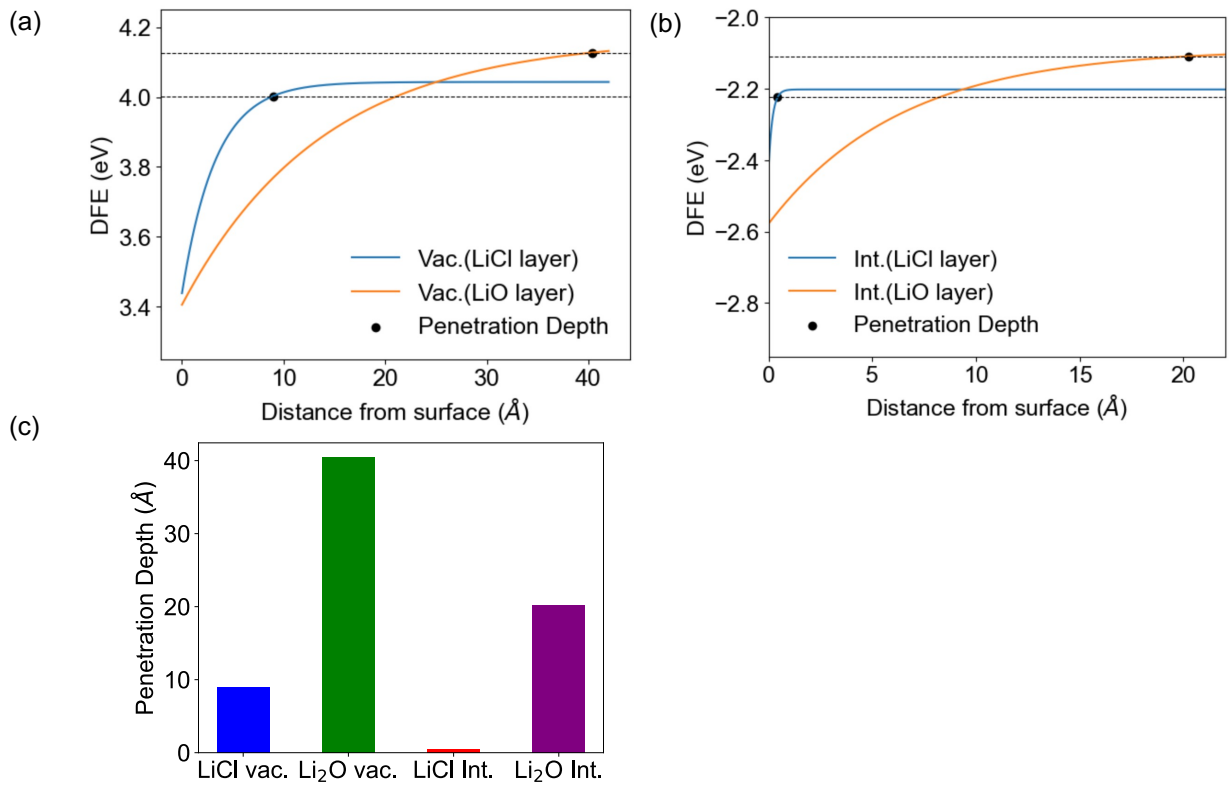

Figure S4: The intersection of the fitted equations with their respective maximum value's 99%. (a) Vacancy in LiCl and Li$_2$O layers, (b) Interstitial in LiCl and Li$_2$O layers, (c) Bar chart showing the penetration depths.

## Relative Structural Distortion:

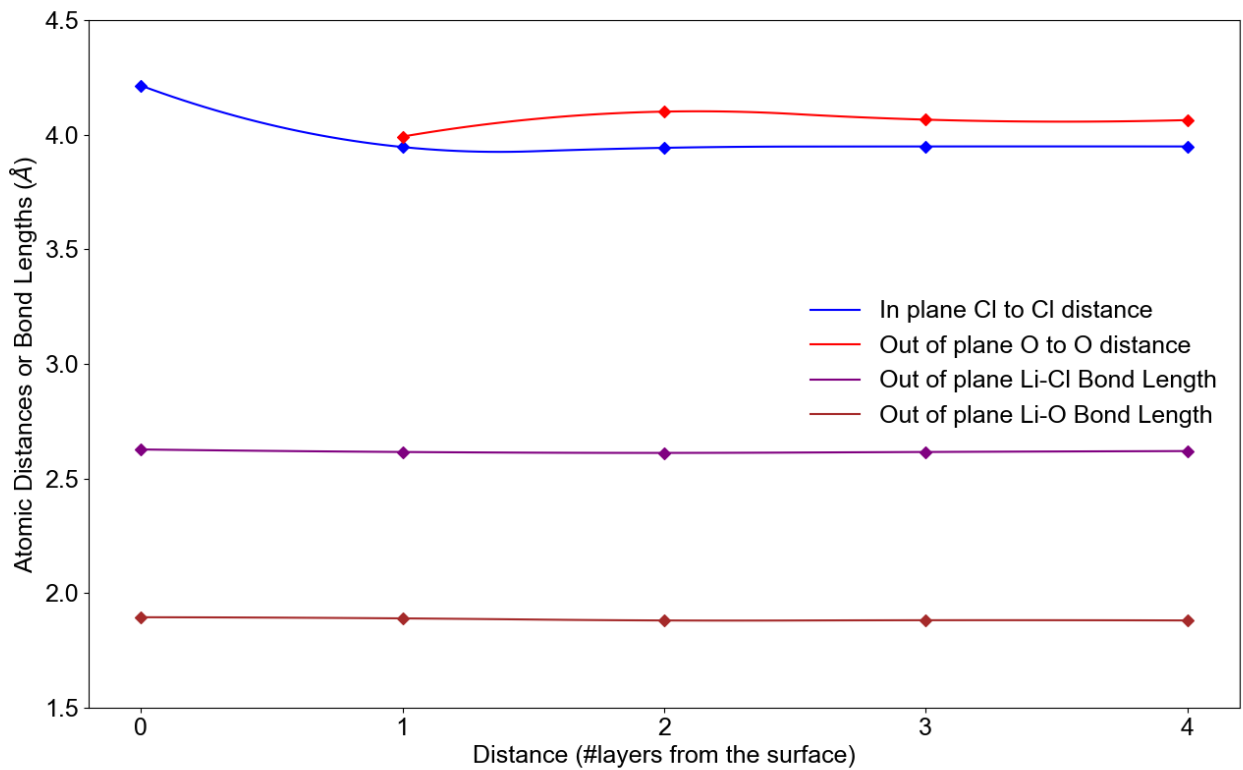

Figure S5: Structural distortion effects resulting from geometric relaxation in a LiCl-terminated 4 by 4 slab. In plane Cl to Cl distance, out of plane O to O distance, out of plane Li-Cl bond length within the LiCl layer containing a vacancy, and out of plane Li-O bond length between the next $Li_2O$ layer and the LiCl layer, plotted as functions of layer number from the surface. Based on the relative comparison of these distances and bond lengths as a function of layer number, it is observed that only in plane Cl to Cl distances appear dominant and play a crucial role in lowering the surface DFE compared to the bulk DFE.

# DFE without relaxation effect:

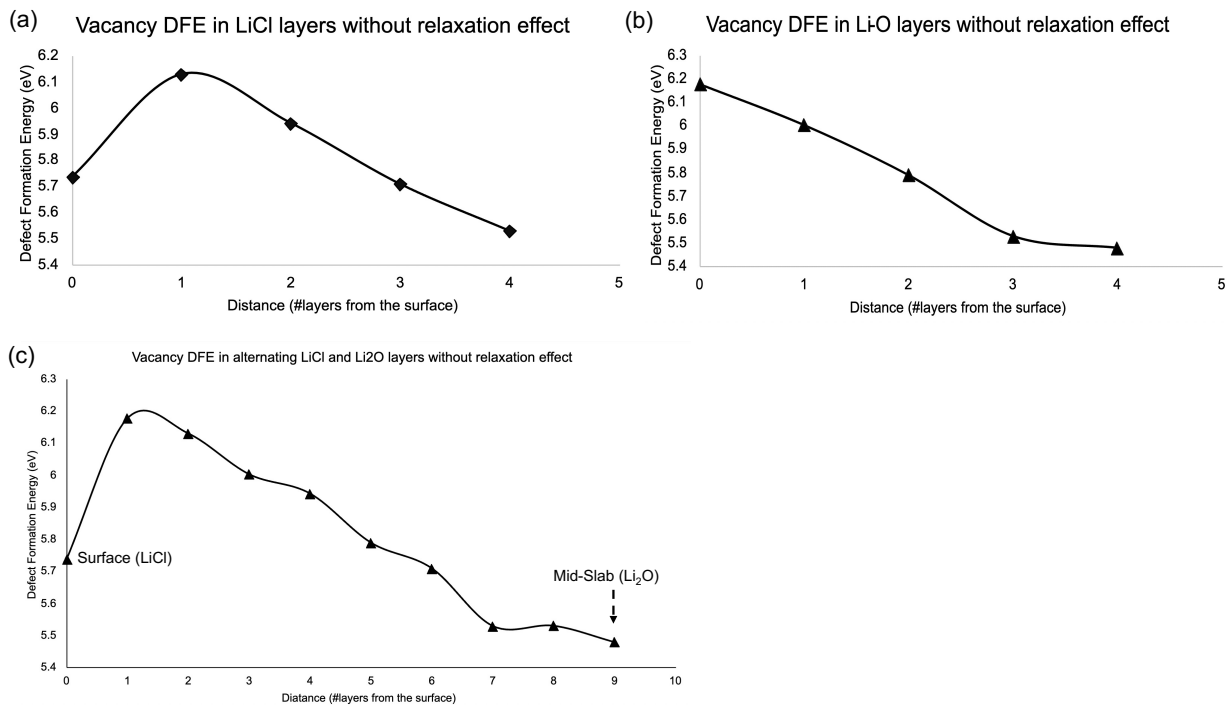

Figure S6: Defect formation energy in a LiCl terminated 4 by 4 supercell slab without the relaxation effect. (a) $Li^+$ Vacancy DFE in LiCl layers, (b) $Li^+$ Vacancy DFE in $Li_2O$ layers, and (c) $Li^+$ Vacancy DFE in alternating LiCl and $Li_2O$ layers. Figure (c) is a combination of (a) and (b).

# Defect Migration Barrier Test in Bulk Supercells

For the NEB calculations in 3x3x3 bulk supercells we used a k-points mesh of 2x2x2 and for the 2x2x2 bulk supercells we used a k-points mesh of 3x3x3. The force convergence criteria used for ionic relaxation was less than 0.05 eV/Å.

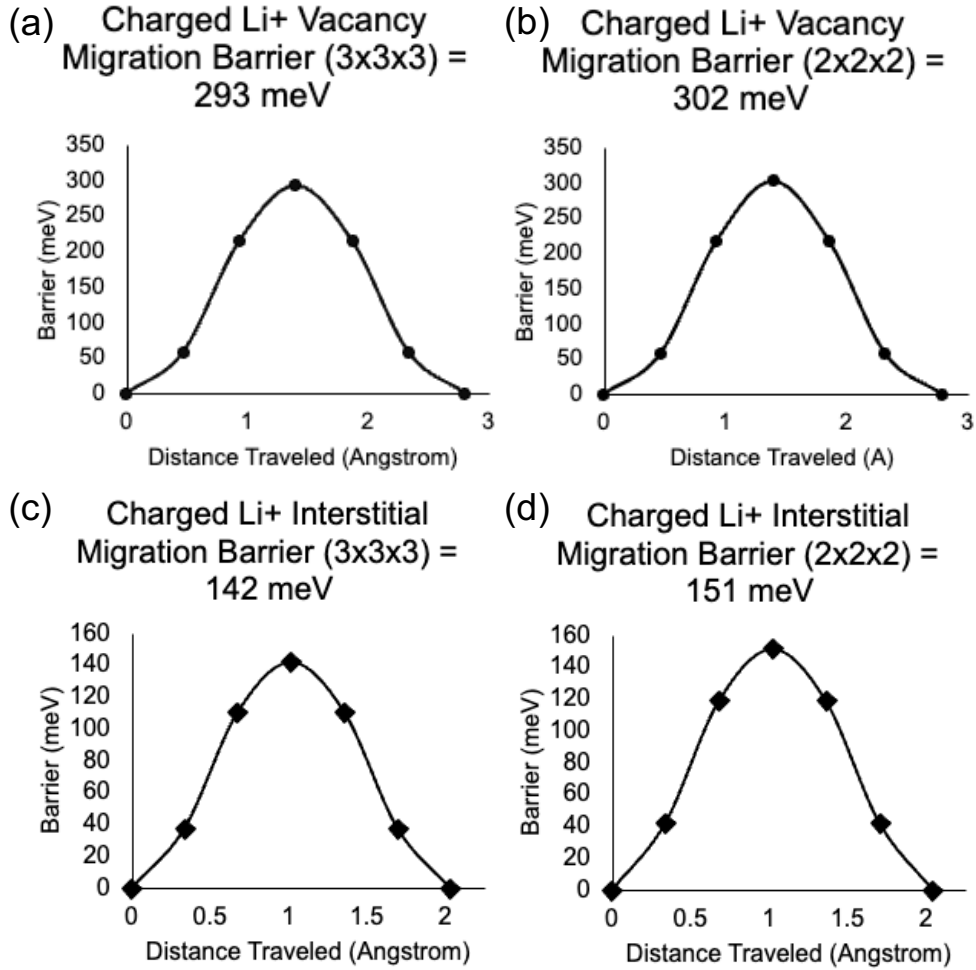

Figure S7: Charged Li$^+$ vacancy migration barrier in (a) 3x3x3 supercell and (b) 2x2x2 supercell. Charged Li$^+$ interstitial dumbbell migration barrier in (c) 3x3x3 supercell and (d) 2x2x2 supercell.


\end{document}